\documentclass[twocolumn,prb,reprint,superscriptaddress,floatfix,amsmath,amssymb,aps]{revtex4-2}
\pdfoutput=1

\usepackage[utf8]{inputenc}
\usepackage[caption=false,subrefformat=parens,labelformat=parens]{subfig}
\usepackage{graphicx}
\usepackage{float}
\usepackage{amsmath}
\usepackage{dcolumn}
\usepackage{bm}
\usepackage{xcolor}
\usepackage{xfrac}
\usepackage[colorlinks=true, allcolors=blue]{hyperref}
\usepackage{chemformula}
\usepackage{siunitx}
\usepackage[normalem]{ulem}

\begin{document}

\preprint{APS/123-QED}

\title{High pressure study of low-Z superconductor \ch{Be22Re}}

\author{J.\ Lim}
\affiliation{Department of Physics, University of Florida, Gainesville, Florida 32611, USA}
\author{A.\ C.\ Hire}
\affiliation{Department of Materials Science and  Engineering, University of Florida, Gainesville, Florida 32611, USA}
\affiliation{Quantum Theory Project, University of Florida, Gainesville, Florida 32611, USA}
\author{Y.\ Quan}
\affiliation{Department of Physics, University of Florida, Gainesville, Florida 32611, USA}
\affiliation{Department of Materials Science and  Engineering, University of Florida, Gainesville, Florida 32611, USA}
\affiliation{Quantum Theory Project, University of Florida, Gainesville, Florida 32611, USA}
\author{J.\ Kim}
\affiliation{Department of Physics, University of Florida, Gainesville, Florida 32611, USA}
\author{L.\ Fanfarillo}
\affiliation{Department of Physics, University of Florida, Gainesville, Florida 32611, USA}
\affiliation{Scuola Internazionale Superiore di Studi Avanzati (SISSA), Via Bonomea 265, 34136 Trieste, Italy}
\author{S.\ R.\ Xie}
\affiliation{Department of Materials Science and  Engineering, University of Florida, Gainesville, Florida 32611, USA}
\affiliation{Quantum Theory Project, University of Florida, Gainesville, Florida 32611, USA}
\author{R.\ S.\ Kumar}
\affiliation{Department of Physics, University of Illinois at Chicago, Chicago, Illinois 60607, USA}
\author{C.\ Park}
\affiliation{HPCAT, X-ray Science Division, Argonne National Laboratory, Argonne, Illinois 60439, USA}
\author{R.~J.~Hemley}
\affiliation{Department of Physics, University of Illinois at Chicago, Chicago, Illinois 60607, USA}
\affiliation{Department of Chemistry, University of Illinois at Chicago, Chicago, Illinois 60607, USA}
\author{Y.\ K.\ Vohra}
\affiliation{Department of Physics, University of Alabama at Birmingham, Birmingham, Alabama 35294, USA}
\author{R.\ G.\ Hennig}
\affiliation{Department of Materials Science and  Engineering, University of Florida, Gainesville, Florida 32611, USA}
\affiliation{Quantum Theory Project, University of Florida, Gainesville, Florida 32611, USA}
\author{P.\ J.\ Hirschfeld}
\affiliation{Department of Physics, University of Florida, Gainesville, Florida 32611, USA}
\author{G.\ R.\ Stewart}
\affiliation{Department of Physics, University of Florida, Gainesville, Florida 32611, USA}
\author{J.\ J.\ Hamlin}
\affiliation{Department of Physics, University of Florida, Gainesville, Florida 32611, USA}

\date{\today}

\begin{abstract}
With $T_c \sim \SI{9.6}{K}$, \ch{Be22Re} exhibits one of the highest critical temperatures among Be-rich compounds.
We have carried out a series of high-pressure electrical resistivity measurements on this compound to \SI{30}{GPa}.
The data show that the critical temperature $T_c$ is suppressed gradually at a rate of $dT_c/dP = \SI{-0.05}{K/GPa}$.
Using density functional theory (DFT) calculations of the electronic and phonon density of states (DOS) and the measured critical temperature, we estimate that the rapid increase in lattice stiffening in Be$_{22}$Re overwhelms a moderate increase in the electron-ion interaction with pressure, resulting in the decrease in $T_c$.
High pressure x-ray diffraction measurements show that the ambient pressure crystal structure of \ch{Be22Re} persists to at least \SI{154}{GPa}.
We discuss the relationship between low-Z Be-rich superconductors and the high-$T_c$ superhydrides.
\end{abstract}

\maketitle

\section{Introduction}
In 1968, Ashcroft proposed that a metallic form of dense hydrogen would become a high-temperature superconductor (HTSC) within the framework of Bardeen-Cooper-Schrieffer (BCS) theory~\cite{Ashcroft_MetallicHydrogen_1968}.
The high transition temperature derives in part from the uniquely high vibrational phonon frequencies associated with the lightest of all elements, hydrogen.
The experimental realization of metallic hydrogen, however, has been extremely challenging due to the required ultra high pressures - theoretical predictions suggest pressures over $\sim \SI{450}{GPa}$ may be required~\cite{McMinis_MetallicHydrogenTheory_2015}.
Some forty years after the prediction of monatomic metallic hydrogen, Ashcroft~\cite{Ashcroft_HydrogenRich_2004} proposed that hydrogen-rich metallic compounds are also candidates for HTSC.
It was suggested that by subjecting hydrogen to ``chemical precompression'' through the incorporation of  heavier elements, the pressures required to reach the metallic state could be lowered substantially.

Recently, these ideas have been borne out in dramatic fashion, with the discovery of several hydrogen-rich compounds that become superconducting at record-breaking temperatures~\cite{drozdov_conventional_2015,Somayazulu_LaH10_2019,Drozdov_LaH10_2019,snider_room-temperature_2020}.
A key feature of these superconducting phases is that they appear to contain higher ratios of hydrogen than are stable at ambient pressure.
For example, at ambient pressure, the most hydrogen-rich compound formed by hydrogen and sulfur is \ch{H2S}, but under pressure this compound can decompose into S and \ch{H3S} (the later being a high temperature superconductor~\cite{drozdov_conventional_2015}). 
These become stable only at high pressures (often well above $\sim \SI{100}{GPa}$) and most form following treatment at high temperatures (over $\sim \SI{1000}{K}$ by laser heating)~\cite{Somayazulu_LaH10_2019,Drozdov_LaH10_2019,Kong_YH9_2019}.
Compositions of hydrogen:non-hydrogen higher than 6:1  are referred to as ``superhydrides''~\cite{Geballe_Superhydrides_2018} and several such materials have been found to become superconducting at high temperatures, even approaching room temperature (\textit{e.g.} \ch{LaH10}~\cite{Somayazulu_LaH10_2019,Drozdov_LaH10_2019} and \ch{YH9}~\cite{Kong_YH9_2019}).

Though hydrogen-rich compounds offer a route to extraordinarily high superconducting critical temperatures, the presence of this element produces unique challenges.
The experiments typically start with host materials (lower hydrides) loaded into a diamond cell with dense hydrogen gas.
In addition to the normal challenges of ultra high experiments, the presence of large amounts of highly compressible hydrogen can lead to
a sample that is extremely small ($\sim \SI{10}{\micro\m}$ in diameter),
difficulties establishing good contacts for electrical resistivity measurements,
and an increased likelihood of failure of the diamond anvils.
These challenges make it worthwhile to investigate whether other low-Z (\textit{i.e.}, light element) superconductors may also exhibit high superconducting critical temperatures.

Among systems at ambient pressure, \ch{MgB2} ($T_c = \SI{39}{K}$) provides the best known example of a low-Z superconductor with a high critical temperature~\cite{Nagamatsu_MgB2ambient_2001}. 
The lightest elemental metal, Li ($Z = 3$), exhibits a critical temperature of only \SI{0.4}{mK} at ambient pressure~\cite{Tuoriniemi_LithiumSC0GPa_2007}, but this increases to $15-\SI{20}{K}$ under pressures of $\sim \SI{30}{GPa}$~\cite{Shimizu_DenseLithiumSC_2002,Struzhkin_DenseLithiumSC_2002,Deemyad_LithiumSC67GPa_2003}.
A number of studies have focused on the potential for high-$T_c$ superconductivity in novel lithium based compounds at high pressure (\textit{e.g.}, Ref.~\cite{rosner_prediction_2002}).
Substantially less work has been done on compounds of the second lightest elemental metal, Be ($Z = 4$).
This may be due, in part, to the dangers associated with the inhalation of Be, though alloys and compounds of Be such as Be-Cu are safe to handle and find widespread use.
Elemental Be is a poor superconductor with $T_c = \SI{26}{mK}$~\cite{Falge_Be_1967}, but the potential for high $T_c$ values in Be compounds has been appreciated for some time~\cite{klein_superconductivity_1980}.
Beryllium tends to form compounds that are very Be-rich (\textit{e.g.}, \ch{Be22Re}), but unlike the case of the superhydrides, these low-Z rich compounds can often be synthesized at ambient pressure.
Several Be-rich compounds are found to be superconductors at ambient pressure (\ch{Be13U}~\cite{Ott_UBe13_1983}, \ch{Be13Th}~\cite{ThBe13LuBe13_2018}, \ch{Be13Lu}~\cite{ThBe13LuBe13_2018}, and \ch{Be22Re}), while several others have yet to be reported to be superconducting, (\textit{e.g.}, \ch{Be13La}, \ch{Be13Y}, \ch{Be13Re}).
It is interesting to consider whether higher $T_c$ values can be induced in Be rich compounds through the application of high pressure.

Among Be rich compounds, \ch{Be22Re} displays one of the highest critical temperatures, with $T_c = \SI{9.6}{K}$ (nearly 400 times higher than the $T_c$ of elemental Be~\cite{Falge_Be_1967}).
Superconductivity was reported in the \ch{Be22X} ($X$ = Mo, W, Tc, or Re) family of compounds by Bucher and Palmy in 1967~\cite{bucher_superconductivity_1967}.
The crystal structure is a cubic, \ch{ZrZn22}-type, with space group ${Fd}\overline{3}m$ (No.\ 227) and $Z = 8$ formula units per conventional unit cell.
The structure is reminiscent of the clathrate-like structures found in certain super-hydrides at high pressure~\cite{Salke2019}.
Recent measurements on \ch{Be22Re} indicate isotropic $s$-wave superconductivity~\cite{Shang_ReBe22ambient_2019}.

In this work, we report the pressure-dependence of the superconducting transition temperature in \ch{Be22Re} revealed by high-pressure electrical resistivity measurements to \SI{30}{GPa}.
The superconducting transition temperature, $T_{c}$, decreases monotonically with increasing pressure at a rate of $\SI{-0.055(3)}{K/GPa}$.
Density functional theory (DFT) calculations indicate that pressure decreases the electronic density of states at the Fermi level $N(0)$ , and we present arguments why the effect of pressure on the electron-phonon matrix element and the electronic density of states is weak, such that lattice stiffening dominates the reduction of $T_c$ with pressure.

\section{Methods}
\label{sec:methods}
Polycrystalline \ch{Be22Re} was synthesized by arc-melting.
Be and Re in stoichiometric amounts (with 3\% excess Be added to account for mass loss during melting) were arc melted together three times using 99.5\% pure Be from Brush Wellman and 99.97\% pure Re from Alfa Aesar.

\begin{figure}
    \centering
    \includegraphics[width=\columnwidth]{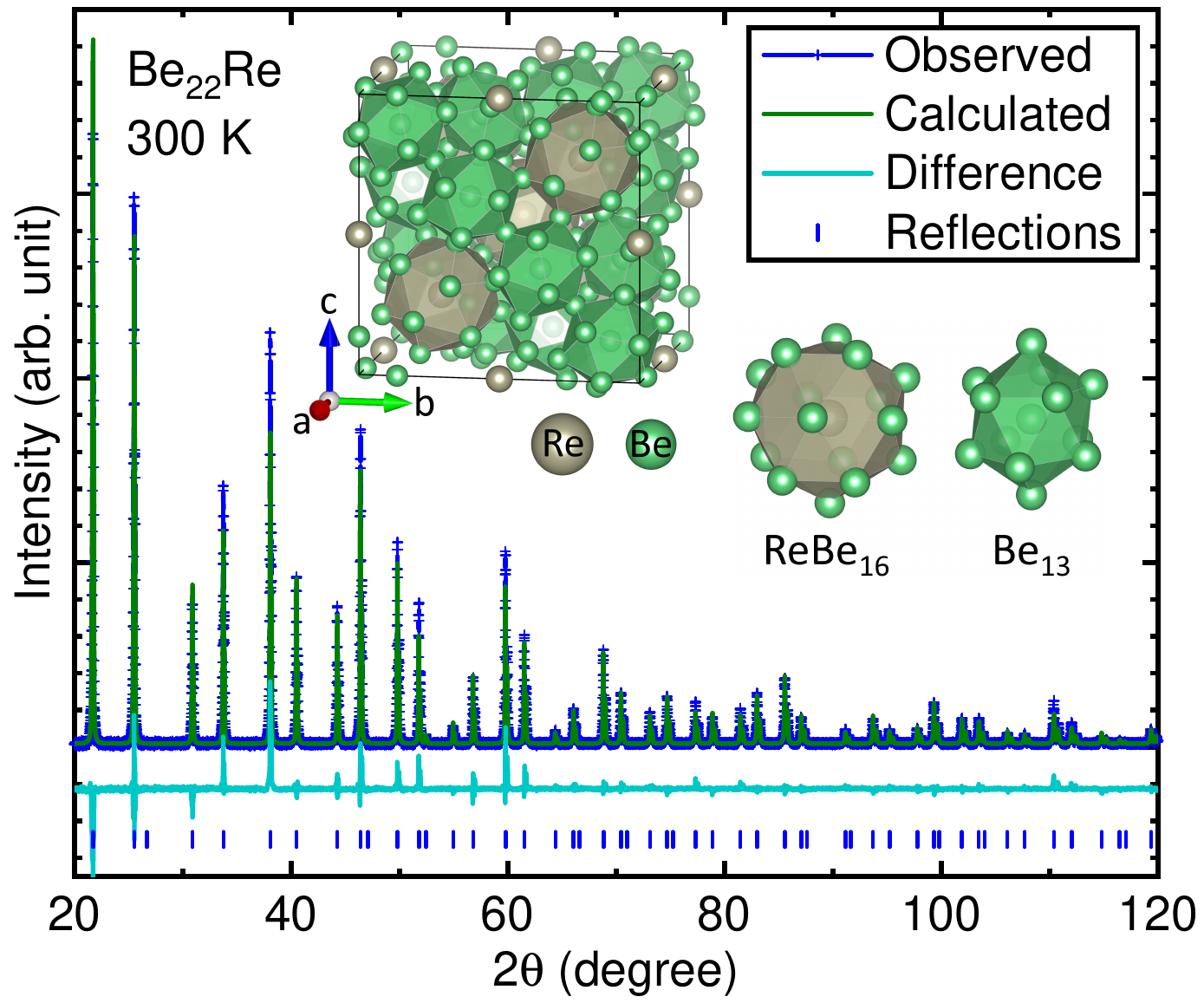}
    \caption{X-ray diffraction pattern of \ch{Be22Re} at ambient pressure. The blue tick marks indicate the expected locations of the peaks.  No additional peaks, indicative of impurity phases, were detected. Inset shows the crystal structure of cubic \ch{Be22Re} (8 formula units) with $a_{0} = \SI{11.568}{\angstrom}$, $V_{0} = \SI{1547.86}{\angstrom^{3}}$, and $\rho_{0} = \SI{3.299}{\gram/\centi\meter^{3}}$ in good agreement with the previous study~\cite{Sands1962}.}
    \label{fig:ReBe22_XRD}
\end{figure}

\begin{figure}
    \centering
    \includegraphics[width=0.8\columnwidth]{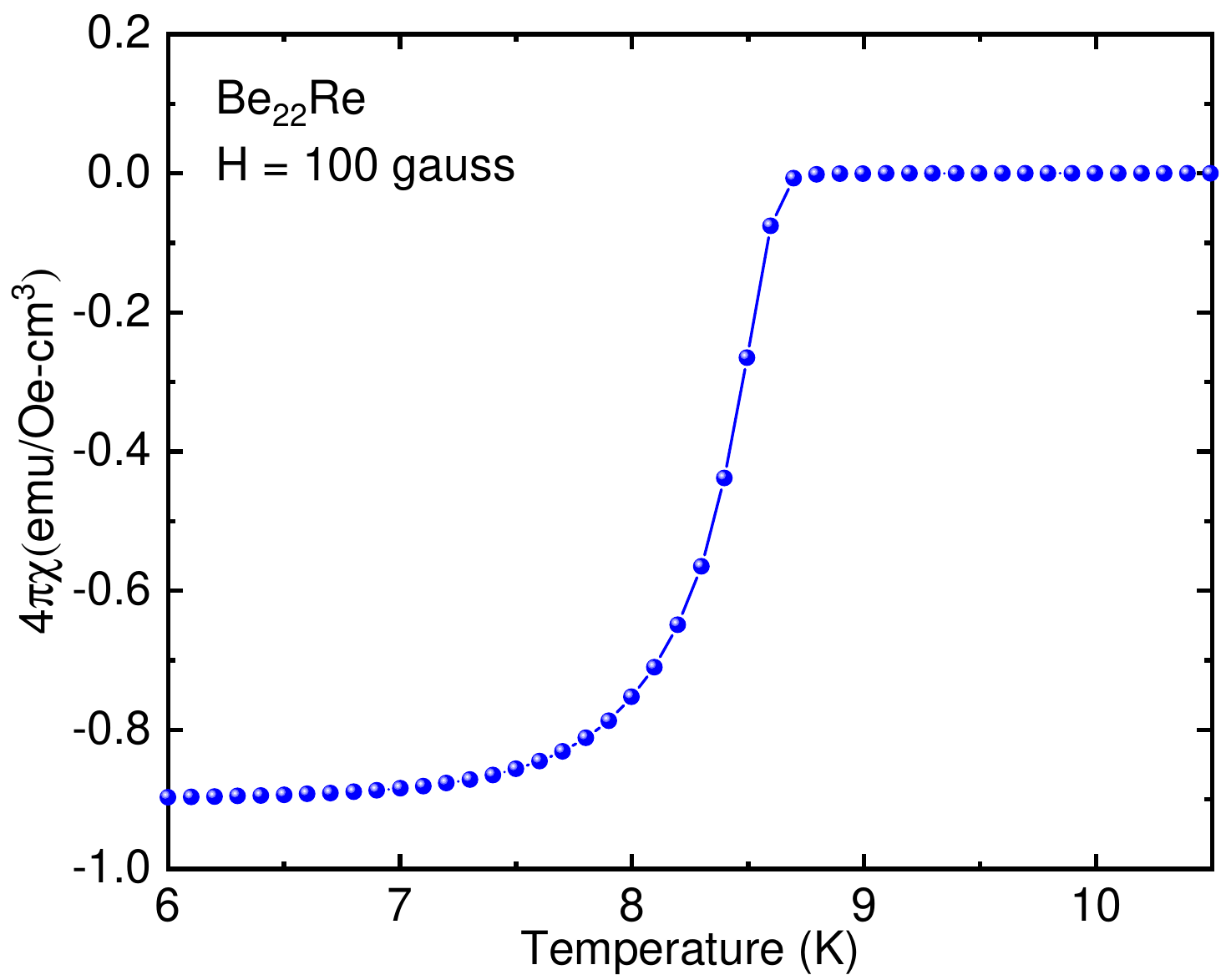}
    \caption{Magnetic susceptibility of \ch{Be22Re} versus temperature at ambient pressure. The data are consistent with full shielding.}
    \label{fig:ReBe22_KiT}
\end{figure}

Powder x-ray diffraction measurements were performed using a Panalytical X'Pert Pro diffractometer. 
Analysis of the diffraction pattern was performed using the software GSAS-II~\cite{toby_gsas-ii_2013} and indicates single phase material after melting (see Fig.~\ref{fig:ReBe22_XRD}).
Magnetic susceptibility measurements performed using a Quantum Design MPMS gave a $T_c$ onset of \SI{8.6}{K} and indicate full shielding (see Fig.~\ref{fig:ReBe22_KiT}).

For the high-pressure resistivity measurements, a micron-sized \ch{Be22Re} polycrystal sample ($\sim$30 $\times$ 30 $\times$ 5 $\mu$m\textsuperscript{3}) was cut from a larger piece of bulk sample and placed in a gas-membrane-driven diamond anvil cell (OmniDAC from Almax-EasyLab) along with a ruby ($\sim$10 $\mu$m in diameter) for pressure calibration~\cite{chijioke_ruby_2005}.
Two opposing diamond anvils (0.15 and 0.5 mm central flats) were used, one of which was a designer-diamond anvil (0.15 mm central flat) with six symmetrically deposited tungsten microprobes in the encapsulated high-quality-homoepitaxial diamond~\cite{weir_epitaxial_2000}.
A 316 stainless steel metal gasket was pre-indented from $\sim$150 to 25 $\mu$m in thickness with a hole ($\sim$80 $\mu$m in diameter), which was filled with soapstone (steatite) for electrically insulating the sample from the gasket and also serving as the pressure-transmitting medium.
The diamond cell was placed inside a customized continuous-flow cryostat (Oxford Instruments).
A home-built optical system attached to the bottom of the cryostat was used for the visual observation of the sample and for the measurement of the ruby manometer.
Pressure was applied at $\sim \SI{8}{K}$ to the desired pressure, and then the sample was cooled down to \SI{5}{K} and warmed up to \SI{15}{K} at a rate of $\sim \SI{0.5}{K/min}$ at each pressure for the temperature-dependent resistivity measurement.
During compression around 8 GPa, pressure was accidentally unloaded to $\sim \SI{3}{GPa}$ and then increased again to 15 GPa.

To estimate the electrical resistivity from the resistance, we used the van der Pauw method, (assuming an isotropic sample in the measurement plane), ${\rho} = {\pi}tR/\ln{2}$, where $t$ is the sample thickness ($\sim\SI{5}{\micro\meter}$) with currents of $0.1-\SI{2}{\milli\ampere}$.
The accuracy of the estimated resistivity is roughly a factor of two considering uncertainties in the initial thickness of the sample.
No attempt was made to take into account the changes in the sample thickness under high pressures.

The high-pressure angle-dispersive X-ray diffraction (ADXRD) experiments on \ch{Be22Re} powder sample were carried out at beamline 16-BM-D, Advanced Photon Source (APS), Argonne National Laboratory. The X-ray beam with a wavelength of $\SI{0.4133}{\angstrom}$ ($\SI{30.00}{\kilo\electronvolt}$) was focused to $\sim\SI{5}{\micro\meter}$ (vertical) $\times$ $\SI{5}{\micro\meter}$ (horizontal) (FWHMs) at the sample position. X-ray diffraction intensities were recorded using a MAR345 image plate detector. The typical exposure time was $\sim$ 60 to 120 seconds per image depending on the sample position. The sample to detector distance was calibrated using a \ch{CeO2} standard. The pressure inside the DAC was determined using an online ruby spectrometer and the Au grains loaded inside the sample chamber. Ne was used as the pressure transmitting medium. The 2D diffraction images were converted to 1D XRD patterns using the DIOPTAS software~\cite{Dioptas2015}, which were then further analyzed by LHPM-Rietica software and Le Bail methods~\cite{LEBAIL1988}.

For evaluating the density of states (DOS) at the Fermi level as a function of pressure, we used density functional theory (DFT) as implemented in VASP~\cite{KRESSE199615,PhysRevB.54.11169}.
The cutoff energy for the plane-wave basis set was set to 520 eV and a $k$-point density of 60 points per \si{\angstrom}\textsuperscript{-1} was used to relax the structures at various pressures.
We used the Perdew-Burke-Ernzerhof (PBE) generalized gradient approximation~\cite{PhysRevLett.77.3865} (GGA) for the exchange-correlation energy along the projector augmented wave (PAW) pseudopotentials~\cite{PhysRevB.50.17953}.
To obtain accurate electronic DOS values, the tetrahedron method with Bl\"ochl correction was used~\cite{PhysRevB.49.16223}.

As the primitive cell contains 46 atoms, a complete calculation of the full phonon dispersion at multiple pressures is computationally prohibitive.
Therefore, we attempt here to develop a qualitative understanding of trends in $T_c$ with pressure based on estimates using phonon frequencies only at the $\Gamma$ point.
The phonon frequencies are calculated using the finite-difference method~\cite{Kresse_1995_finite_difference}.
To estimate the phonon density of states (PDOS), $F(\omega)$ from the $\Gamma$ point phonon frequencies, we apply a Gaussian smearing with a width of $\sigma = $\SI{1}{THz} (\SI{4.135}{meV}),
\begin{equation}
    F(\omega) = \sum_{i=1}^{N} \frac{D(\omega_i)}{\sqrt{2\pi}\sigma}\exp{\left(-\frac{(\omega-\omega_i)^2}{2\sigma^2}\right)},
    \label{phdos}
\end{equation}
where the summation is over all the $\Gamma$ point phonons, $D(\omega_i)$ is the degeneracy of the phonon with frequency $\omega_i$.

Allen \textit{et al.}~\cite{Allen1975} note that for simple materials (\textit{e.g.}, elements) $\alpha^2 F$ is proportional to $F$. For example, the $\alpha^2 F$ and $F$ for Pb have almost identical shapes and only differ in magnitude by a constant factor (see Appendix I of Ref.~\cite{Allen1975}).
Assuming that the same holds also for \ch{Be22Re}, we approximate $\alpha^2 F$ as
\begin{equation}
    \alpha^2 F = k N(0) F(\omega),
    \label{eqn:approx}
\end{equation}
where $k$ is a proportionality factor and $N(0)$ is the density of states at the Fermi level, accounting for possible changes in the number of states that can couple to the phonons. With this approximation, we obtain the Allen-Dynes parameters $\lambda$, $\omega_{\text{log}}$, and $\left<\omega ^2 \right>$~\cite{Allen1975} from the phonon density of states as
\begin{equation}
    \lambda = 2 \int_0^{\infty} d\omega \frac{\alpha^2F(\omega)}{\omega} = 2k N(0)\int_0^{\infty} d\omega \frac{F(\omega)}{\omega},
    \label{eqn:lambda}
\end{equation}
\begin{align}
    \omega_{\text{log}} & = \exp\left[\frac{2}{\lambda}\int_0^{\infty} d\omega \frac{\alpha^2F(\omega)}{\omega} \ln{\omega}\right] \nonumber \\
    & = \exp\left[\frac{2kN(0)}{\lambda}\int_0^{\infty} d\omega \frac{F(\omega)}{\omega} \ln{\omega}\right],
\end{align}
\begin{align}
    \langle \omega^n \rangle &= \frac{2}{\lambda} \int_0^{\infty} d\omega~\alpha^2F(\omega) \omega^{n-1} \nonumber\\
    &= \frac{2kN(0)}{\lambda} \int_0^{\infty} d\omega~F(\omega) \omega^{n-1}.
\end{align}
If $k$ were a function of $\omega$, it would  quantify the interaction/coupling strength between electrons at $E_f$ and the phonons at frequency $\omega$.
In this study, we first assume that $k$ is independent of $\omega$ and pressure.
We then fix the value of the constant $k$ such that the simplified Allen-Dynes equation (\textit{i.e.}, $f_1=f_2=1$)~\cite{Allen1975},
\begin{equation}
        T_c = \frac{\omega_{\text{log}}}{1.20}\exp\left[-\frac{1.04(1+\lambda)}{\lambda-\mu^*(1+0.62\lambda)}\right]
	\label{eqn:AD}
\end{equation}
reproduces the experimental value of $T_c$ at ambient pressure.
In order to obtain trends in $T_c$ that are relevant to the high pressure data, we have used the extrapolated zero pressure $T_c$ value from our high pressure data, $T_c(P=0) \sim \SI{8}{K}$, which is somewhat lower than the ambient pressure $T_c \sim \SI{8.6}{K}$ from susceptibility data on a different piece of sample.
This may be due to strain induced disorder in the sample due to the non-hydrostatic conditions present in the pressure chamber.
The Coulomb pseudopotential, $\mu ^*$, is  approximated as $\mu ^* = 0.1$.
If we instead allow $k$ to vary with pressure such that the experimentally observed $T_c$ values are reproduced, we find that the value of $k$ varies by only about 4\% between ambient pressure and \SI{30}{GPa}.
This suggests that a pressure independent $k$ is a reasonably good approximation.

\section{Results}
\begin{figure}
    \centering
    \includegraphics[width=0.9\columnwidth]{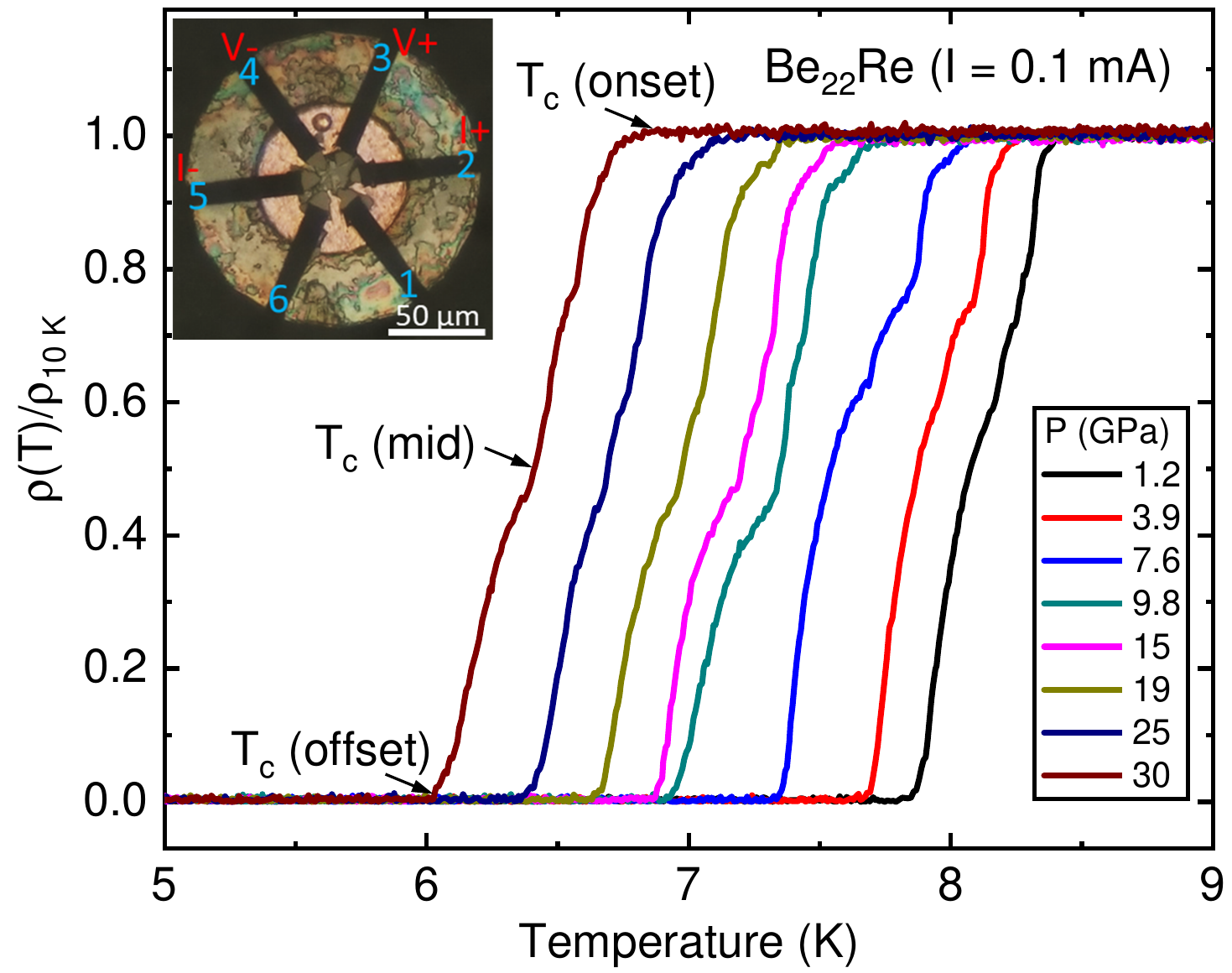}
    \caption{Relative resistivity versus temperature measured while warming at several pressures to \SI{30}{GPa}. Three arrows represent $T_{c}$ (onset), $T_{c}$ (mid), and $T_{c}$ ($\rho = 0$), respectively, as defined in the text. All the data were taken during compression except for those at \SI{9.8}{GPa} which were measured during decompression. The inset shows the photograph of the \ch{Be22Re} sample along with a ruby for pressure calibration, steatite insulation (bright area surrounding the sample at center), a 316 stainless steel metal gasket, and six tungsten leads configuration. Leads 2, 3, 4, and 5 were used for the measurement.}
    \label{fig:ReBe22_relRTP0.1mA}
\end{figure}

Figure~\ref{fig:ReBe22_relRTP0.1mA} shows the relative resistivity $\rho(T)/\rho_{\SI{10}{K}}$ versus temperature for \ch{Be22Re} to pressures of 30 GPa, focusing on the low temperature region near the superconducting transition.
All the resistivity curves are based on compression except for \SI{9.8}{GPa}, which was measured during  decompression.
We define $T_{c}$ (onset), $T_{c}$ (mid), and $T_{c}$ ($\rho = 0$) as the temperatures where the resistivity just begins to drop below the normal state trend, drops to 50\% of the normal state value, and drops to 0, respectively.
The critical temperature $T_{c}$ monotonically decreases with increasing pressure from \SI{8.07}{K} at \SI{1.2}{GPa} to \SI{6.41}{K} at \SI{30}{GPa}.
Ambient pressure resistivity measurements found a transition width of \SI{0.23}{K}~\cite{Shang_ReBe22ambient_2019}.
Under pressure, the width of the superconducting transition, $\Delta T_{c}$, defined as the difference between $T_{c}$ (onset) and $T_{c}$ ($\rho = 0$), increases from \SI{0.6}{K} at \SI{1.2}{GPa} to \SI{0.8}{K} at \SI{30}{GPa}.
The relatively small increase in transition width under pressure indicates that the tiny sample is subject to only small pressure gradients.

\begin{figure}
    \centering
    \includegraphics[width=0.9\columnwidth]{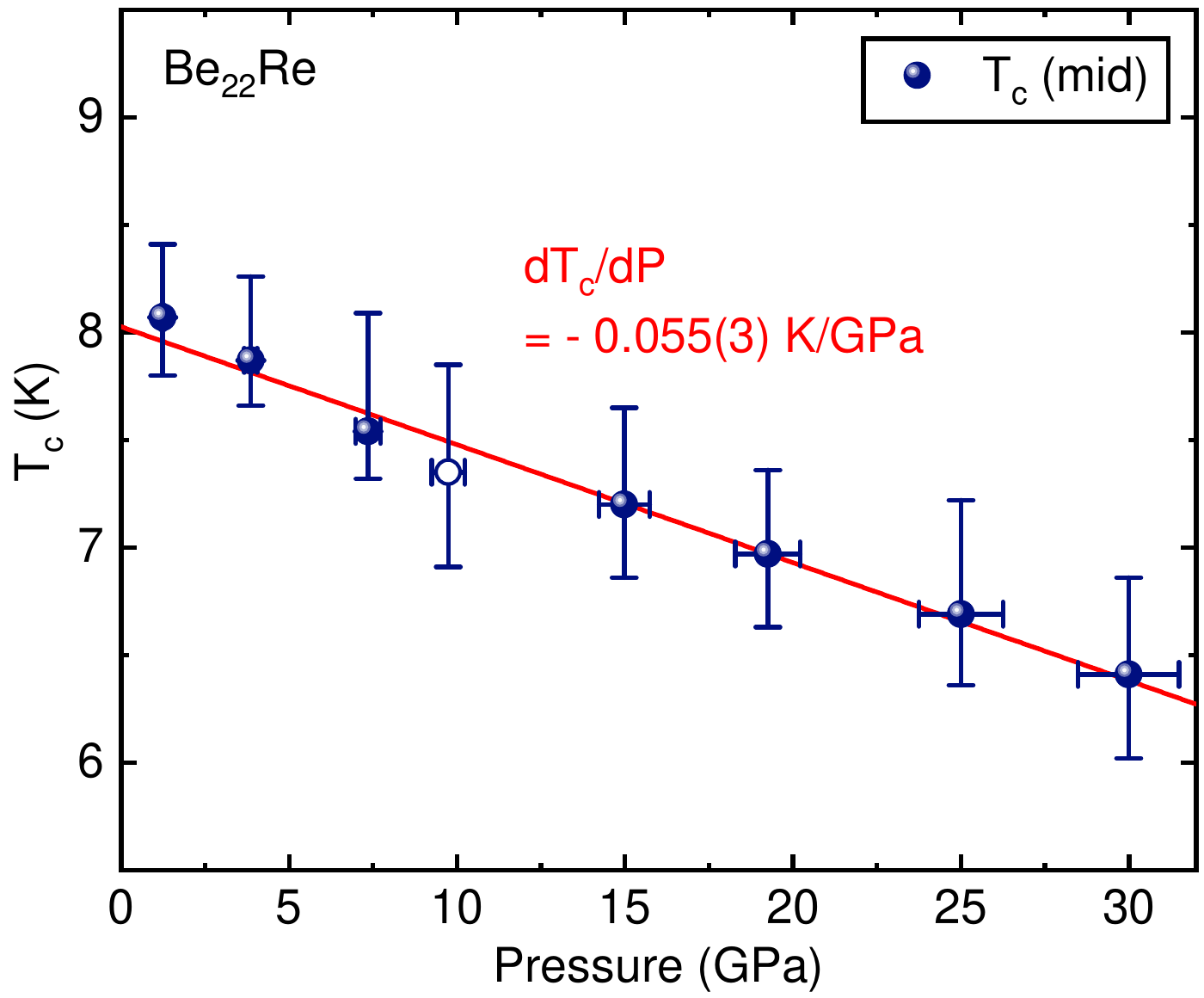}
    \caption{Superconducting transition temperature ($T_{c}$) of \ch{Be22Re} versus pressure. Blue sphere (or open) symbols indicate the midpoint of the transition taken from compression (or decompression). The red solid line refers to the linear fit of $T_{c}$.}
    \label{fig:ReBe22_TcP}
\end{figure}

Figure~\ref{fig:ReBe22_TcP} presents $T_{c}$ versus pressure for \ch{Be22Re} to \SI{30}{GPa}.
The points are taken from the midpoint of the transition while the vertical error bars indicate $T_c$ (onset) and $T_c$ ($\rho = 0$) as defined in Fig.~\ref{fig:ReBe22_relRTP0.1mA}.
The red line represents a linear fit to the midpoint of the transition, which produces a slope \SI{-0.055(3)}{K/GPa}.
The trend is reversible as the data at \SI{9.8}{GPa}, which was measured during decompression, fits well within the trend.
An estimate of $T_{c}$ at ambient pressure from the linear fit yields $\sim \SI{8}{K}$.
This is somewhat lower that the ambient pressure resistivity onset reported by Shang~\textit{et al.}~\cite{Shang_ReBe22ambient_2019}, but is consistent with the midpoint of the susceptibility transition that we measured (see Fig.~\ref{fig:ReBe22_KiT}).

\begin{figure}
    \centering
    \includegraphics[width=0.8\columnwidth]{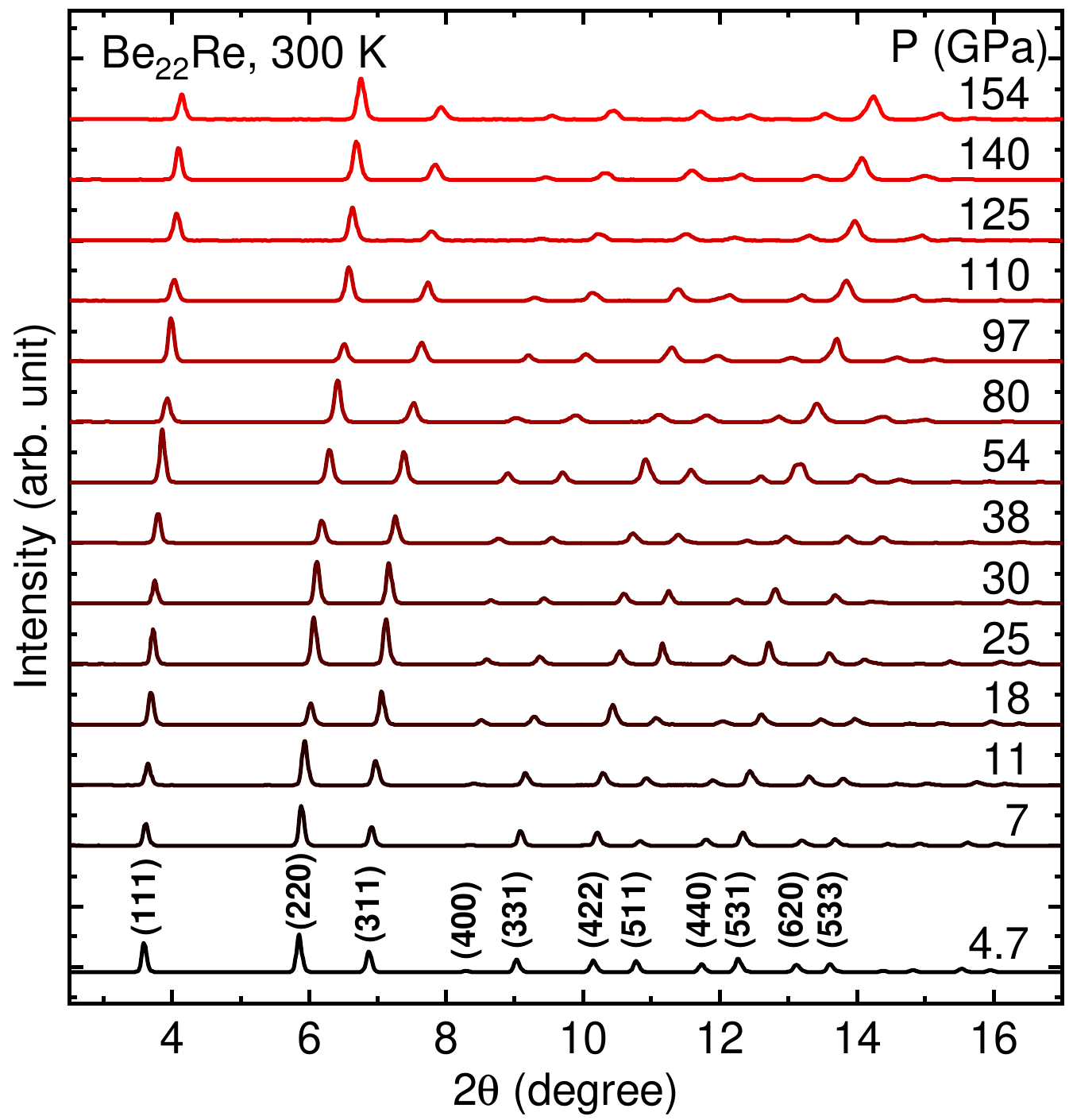}
    \caption{Representative high-pressure XRD patterns of \ch{Be22Re} at pressures to 154 GPa. No structural transition was observed throughout the pressure range studied.}
    \label{fig:ReBe22_HPXRD}
\end{figure}

\begin{figure}
    \centering
    \includegraphics[width=0.9\columnwidth]{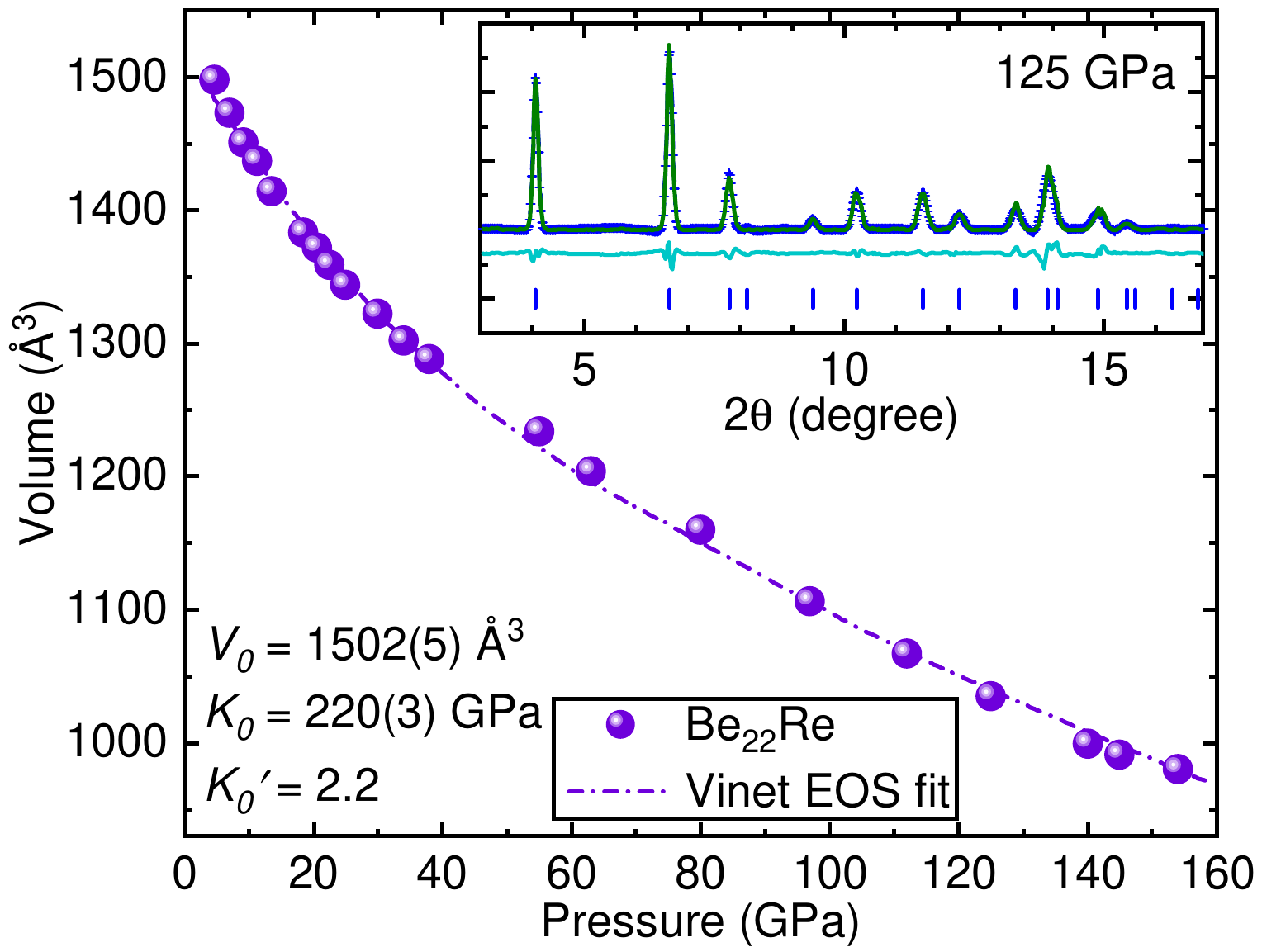}
    \caption{PV-isotherm of \ch{Be22Re} at room temperature. Inset shows the Le Bail fit at 125 GPa with the initial cubic structure.}
    \label{fig:ReBe22_PVLeBail}
\end{figure}

High-pressure X-ray diffraction patterns measured at room temperature to pressures as high as \SI{154}{GPa} are shown in Fig.~\ref{fig:ReBe22_HPXRD}.
Clearly, no structural transition is observed throughout the pressure range underscoring the significant stability of the initial cubic structure of \ch{Be22Re}.
Some XRD patterns, for example at 18 and 97 GPa, show the presence of preferred orientation depending on the sample position, which is introduced by the nonhydrostatic pressure condition.
The resulting pressure-volume (PV) curve is shown in Fig.~\ref{fig:ReBe22_PVLeBail}, which is fitted with Vinet equation of state (EOS)~\cite{Vinet1987}.
The fit produces the relatively high value of bulk modulus ($K_0$), 220 GPa for \ch{Be22Re}, compared to that of Be metal, 114 GPa~\cite{Lazichi2012}.
It is clear that despite the low concentration, the dilute Re plays an important role for the hardness of \ch{Be22Re} given that the bulk modulus of Re metal is \SI{353}{GPa}~\cite{Anzellini2014}.
The inset of Fig.~\ref{fig:ReBe22_PVLeBail} shows the refined diffraction pattern at 125 GPa in terms of the initial cubic structure using the Le Bail method~\cite{LEBAIL1988}, which confirms the absence of any structural transition.

\begin{figure}
    \centering
    \captionsetup{justification=centering}
    \includegraphics[width=0.9\columnwidth]{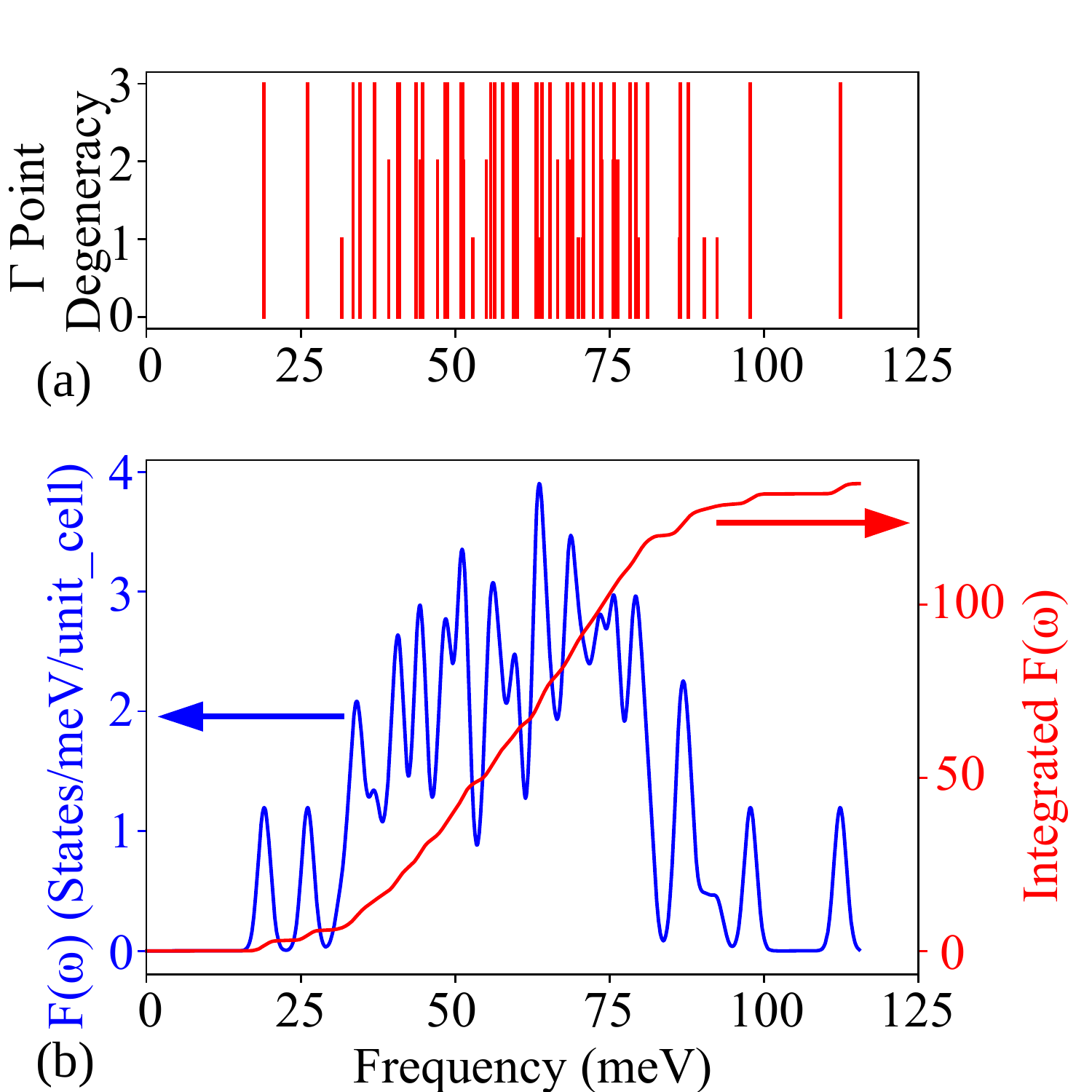}\\
    \caption{(a) $\Gamma$ point phonon frequencies and their respective degeneracies at 0~GPa (b) Calculated phonon DOS given a smearing factor $\sigma= 4.135$~meV (\SI{1}{THz}).}
    \label{fig:dos_press}
    \label{ph_occ}
\end{figure}

\begin{figure}
    \centering
    \captionsetup{justification=centering}
    \includegraphics[width=0.80\columnwidth]{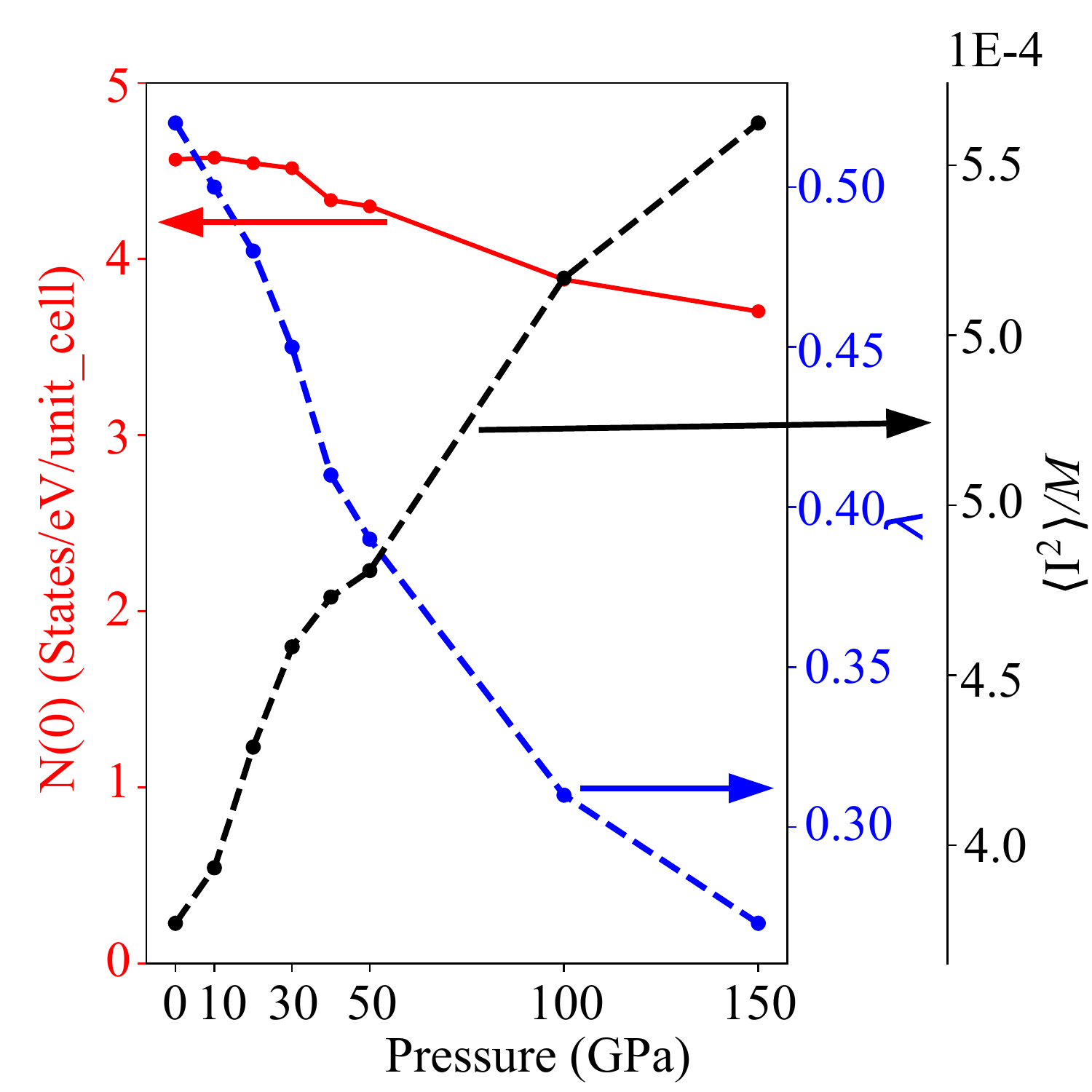}
    \caption{Calculated density of states at the Fermi level $N(0)$ (red curve - both spins are included), electron-phonon coupling parameter $\lambda$ (blue curve) and  $\langle I^2 \rangle/M$ (black curve) as a function of pressures.}
    \label{dos_lambda}
\end{figure}

Figure~\ref{ph_occ}(a) and (b) shows the VASP calculated $\Gamma$ point phonons along with their degeneracies and the calculated phonon density of states using Gaussian smearing as described in  Sec.~\ref{sec:methods}.
Increasing the smearing $\sigma$ leads to the smoothing of the phonon density of states. 
The resulting values of the integrated $F(\omega)$ are not strongly dependent on the smearing for reasonable values of $\sigma$.
Figure~\ref{dos_lambda} shows the pressure dependence of the density of states at the Fermi level $N(0)$ (red curve) and the electron-phonon coupling parameter $\lambda$ (blue curve).
$\lambda$ is determined using Eqn.~\eqref{eqn:lambda}, with the value of the constant $k$ determined using the extrapolated zero pressure $T_c = \SI{8}{K}$ from the high pressure data.
$N(0)$ slightly monotonically decreases with pressure by about 20\%, while $\lambda$ decreases monotonically by nearly a factor of two between ambient pressure and \SI{150}{GPa}.

The electron-phonon coupling parameter $\lambda$ can also be represented by,
\begin{equation}
    \lambda = \dfrac{N(0)\left< I^2 \right>}{M \left< \omega ^2 \right >}
    = \dfrac{\eta}{M\left< \omega ^2 \right >},
    \label{eqn:lambda_i2}
\end{equation}
where $\left< I^2 \right>$ is the Fermi surface averaged electron-phonon matrix element, $M$ represents an average atomic mass, and $\eta$ is the McMillan-Hopfield parameter~\cite{mcmillan_transition_1968,hopfield_angular_1969,hopfield_systematics_1971}.
From Eqn.~\eqref{eqn:lambda_i2} we can extract the value of $\left< I^2 \right>/M$ as a function of pressure and this is plotted as the black data points in Fig.~\ref{dos_lambda}.
Allen and Dynes~\cite{Allen1975} have highlighted that $\eta$ is one of the key parameters controlling the superconducting critical temperature.
We find that $\eta$ increases by more than 50\% from ambient pressure to \SI{30}{GPa}.

\begin{figure}
    \centering
    \captionsetup{justification=centering}
    \includegraphics[width=0.95\columnwidth]{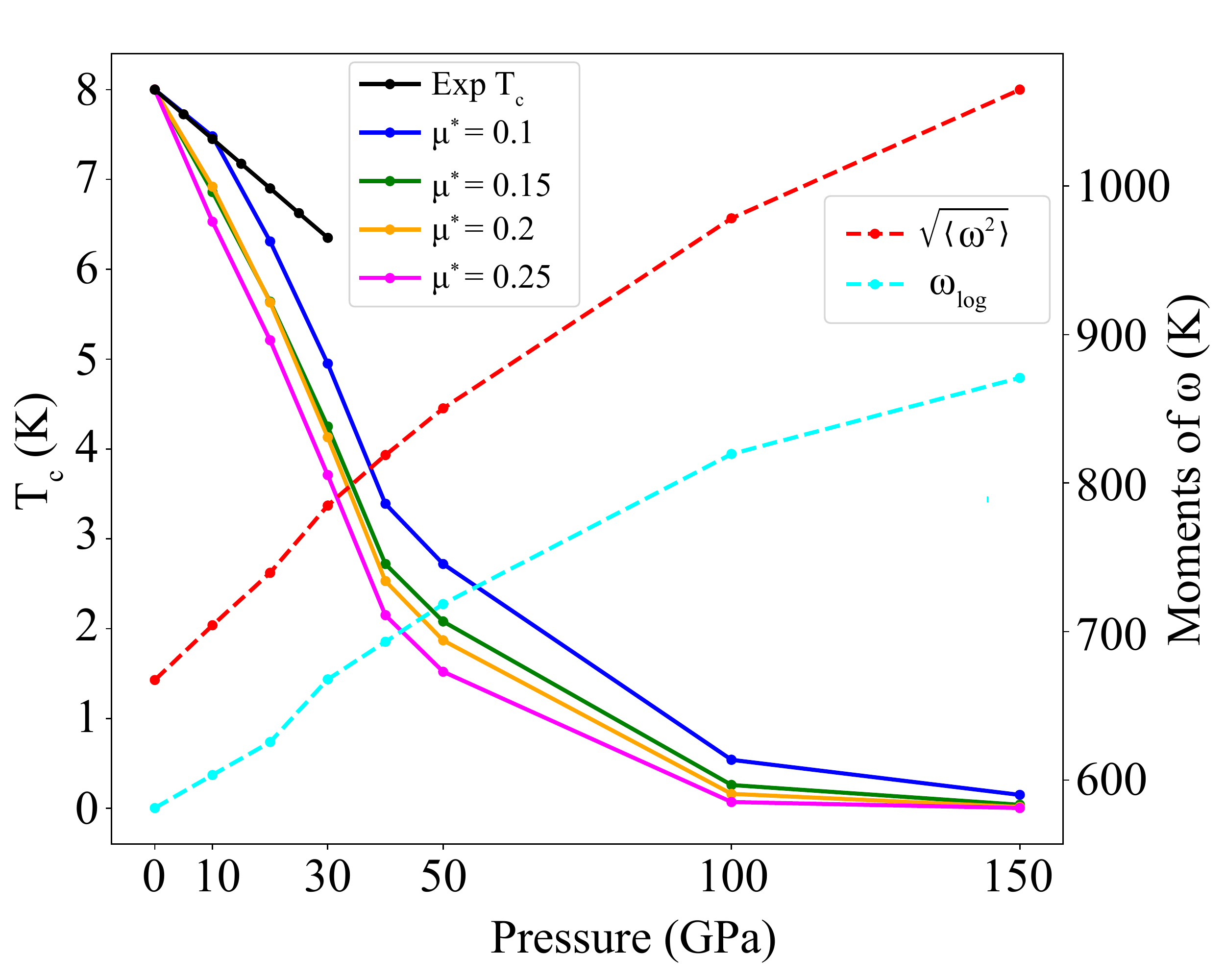}
    \caption{Experimental and calculated $T_c$, and calculated $\sqrt{\langle \omega^2 \rangle}$ and $\omega_{\text{log}}$ as a function of pressure. The $T_c$ was calculated using Allen-Dynes equation at 4 values of $\mu^*$.}
    \label{tc_mu_star}
\end{figure}

By combining the information obtained on the density of states at the Fermi level, electron-phonon coupling parameter, and phonon frequencies, we can estimate the expected pressure dependence of $T_c$.
Figure~\ref{tc_mu_star} illustrates the pressure dependence of the phonon frequencies (dashed curves), experimental $T_c$ (black curve), and the computationally estimated trend in $T_c$ based on the approximation represented by Eqn.~\eqref{eqn:approx} together with the Allen-Dynes equation (Eqn.~\ref{eqn:AD}).
Estimates are provided for different values of the Coulomb pseudopotential $\mu ^*$ and, as expected, the values of $T_c$ depend rather weakly on this parameter.
In the region up to \SI{30}{GPa} where experimental data exist, the agreement is reasonably good, with $T_c$ underestimated by only 25\% at \SI{30}{GPa} for $\mu ^* = 0.1$.
Since we know from the high pressure x-ray diffraction data that the crystal structure remains unchanged to at least \SI{150}{GPa}, we can use this method to estimate $T_c$ to pressures beyond the range of the resistivity experiments.
Based on the calculations we find that $T_c$ continues to decrease, reaching a value below \SI{1}{K} at \SI{150}{GPa}.

Changes in $\lambda$ (and consequently $T_c$) are controlled by the relative changes in $\eta$ and $\langle \omega ^2 \rangle$.
The observed decrease in $\lambda$ with pressure can be understood as deriving from the fact that lattice stiffening (increase in $\langle \omega ^2 \rangle$) dominates over electronic effects (increase in $\eta$).
At low pressures, we find that the logarithmic volume derivative of $\eta$ takes on a value of $d\ln{\eta}/d\ln{V} \approx -1.2$.
This value is similar to that found for many simple metal ($s$, $p$) superconductors (including \textit{e.g.}, \ch{MgB2}) and is significantly smaller than the value of $\sim -3.5$ found in many transition metals~\cite{tomita_dependence_2001}.
The comparatively small magnitude of $d\ln{\eta}/d\ln{V}$ in elemental simple metals causes $T_c$ to decrease with pressure initially~\cite{hamlin_superconductivity_2015}.
The fact that \ch{Be22Re} behaves as a simple metal in regards to superconductivity under pressure is consistent with the fact that the $N(0)$ is dominated by Be $2p$ electrons~\cite{Shang_ReBe22ambient_2019}.

Adjacent to Be in the periodic table, Li is a prototypical simple metal at ambient pressure, but exhibits a remarkable divergence from simple metal behavior at high pressure.
Under pressure, Li becomes superconducting at temperatures approaching \SI{20}{K}, exhibits complex crystal structures, and even becomes semiconducting above \SI{75}{GPa}~\cite{neaton_pairing_1999,hanfland_new_2000,Shimizu_DenseLithiumSC_2002,Struzhkin_DenseLithiumSC_2002,Deemyad_LithiumSC67GPa_2003,matsuoka_direct_2009,lv_predicted_2011}.
The anomalous behavior of Li has been attributed to the influence of the ion cores, which approach each other at high pressure, increasingly restrict the valence electrons to low symmetry interstitial regions, and eventually localize them enough to produce semiconducting behavior~\cite{neaton_pairing_1999,lv_predicted_2011,matsuoka_direct_2009}.
Similar physics is thought to influence the behavior of certain Li-rich compounds which have either been found ~\cite{matsuoka_pressure-induced_2008} to exhibit superconductivity ($T_c$ = \SI{13}{K}) under pressure or have been predicted to exhibit higher temperature superconductivity or complex crystal structures under pressure~\cite{neaton_pairing_1999,feng_theoretical_2007,feng_emergent_2008}.
However, the same evolution of complex crystal structures does not appear likely to occur in Be-rich compounds because the ion cores of Be are 25-40\% smaller than those of Li~\cite{waber_orbital_1965,CM_Handbook_2005}.
The size difference is significant enough that even at \SI{300}{GPa}~\cite{lazicki_high-pressure--temperature_2012}, the degree of core overlap for Be is much less than for Li at \SI{75}{GPa} (the pressure where Li becomes semiconducting~\cite{matsuoka_direct_2009,matsuoka_pressure-induced_2014}).
Thus, Be and Be-rich compounds may tend towards simple metal behavior even at multi-megabar pressures.

\section{Conclusions}
In summary, experiments show that the superconducting critical temperature of \ch{Be22Re} is suppressed by pressure to at least \SI{30}{GPa}.
Computational estimates based on electronic density of states and phonon calculations suggest that $T_c$ will continue to be monotonically suppressed at higher pressures. Furthermore, the calculations and measurements indicate that lattice stiffening overcomes electronic effects, leading to the observed decrease in $\lambda$ and $T_c$ with pressure.
High pressure x-ray diffraction shows that the ambient pressure crystal structure is remarkably stable and remains unchanged to at least \SI{150}{GPa}.
This stability is similar to that observed in elemental Be, which remains in the ambient pressure \textit{hcp} structure to at least \SI{170}{GPa}~\cite{mcmahon_high-pressure_2006}.

\section*{Acknowledgments}
We acknowledge enlightening discussion with L.\ Boeri and E.\ Zurek.
We thank S. Tkachev (GSECARS, University of Chicago) for sample gas loading for the x-ray diffraction measurements.  
Work at the University of Florida performed under the auspices of U.S. Department of Energy Basic Energy Sciences under Contract No.\ DE-SC-0020385.
A.H.\ acknowledges the support from the Center for Bright Beams, U.S. National Science Foundation award PHY-1549132.
R.K.\ and R.H.\ acknowledge support from the U.S.\ National Science Foundation (DMR-1933622). X-ray diffraction measurements were performed at HPCAT (Sector 16), Advanced Photon Source (APS), Argonne National Laboratory. HPCAT operations are supported by the DOE-National Nuclear Security Administration (NNSA) Office of Experimental Sciences.
The beamtime was made possible by the Chicago/DOE Alliance Center (CDAC), which is supported by DOE-NNSA (DE-NA0003975).
The Advanced Photon Source is a DOE Office of Science User Facility operated for the DOE Office of Science by Argonne National Laboratory under Contract No. DE-AC02-06CH11357.

\newpage
\bibliography{Be22Re}

\end{document}